\newcommand{\gev}{\, {\rm GeV}}
\newcommand{\tev}{\, {\rm TeV}}

\newcommand{\beq}{\begin{equation}}
\newcommand{\eeq}{\end{equation}}
\newcommand{\bea}{\begin{eqnarray}}
\newcommand{\eea}{\end{eqnarray}}
\newcommand{\seff}{\sin^2\theta_W^{\rm eff}}
\newcommand{\gsim}{\lower.7ex\hbox{$\;\stackrel{\textstyle>}{\sim}\;$}}
\newcommand{\lsim}{\lower.7ex\hbox{$\;\stackrel{\textstyle<}{\sim}\;$}}

\documentclass[aps,prl,twocolumn,preprintnumbers,%
               floatfix,nofootinbib]{revtex4}
\usepackage{graphicx}

\def\stacksymbols #1#2#3#4{\def\theguybelow{#2}
    \def\vp{\lower#3pt}
    \def\sp{\baselineskip0pt\lineskip#4pt}
    \mathrel{\mathpalette\intermediary#1}}

\def\intermediary#1#2{\vp\vbox{\sp
     \everycr={}\tabskip0pt
     \halign{$\mathsurround0pt#1\hfil##\hfil$\crcr#2\crcr
              \theguybelow\crcr}}}

\def\to{\rightarrow}

\begin{document}

%
%

\preprint{Saclay T04/084, MCTP-04-37, ANL-HEP-PR-04-63, EFI-04-23}

\title{First-Order Electroweak Phase Transition in the Standard Model with a Low Cutoff}

\author{Christophe Grojean${}^{a,b}$, G\'eraldine Servant${}^{a,c,d}$,
James D. Wells${}^{b}$}
\affiliation{
${}^{(a)}$ Service de Physique Th\'eorique, CEA Saclay, F91191 Gif-sur-Yvette,
France \\
${}^{(b)}$ MCTP, Department of Physics, University of Michigan, Ann Arbor, 
MI 48109 \\
${}^{(c)}$ High Energy Physics Division, Argonne National Laboratory, Argonne,
IL 60539 \\
${}^{(d)}$ Enrico Fermi Institute, University of Chicago, Chicago, IL 60637}


\begin{abstract}
We study the possibility of a first-order electroweak phase transition
(EWPT) due to a dimension-six operator in the effective Higgs
potential. In contrast with previous attempts to make the EWPT
strongly first-order as required by electroweak baryogenesis, 
we do not rely on large one-loop thermally generated  cubic Higgs interactions.
Instead, we augment the Standard Model (SM) effective theory with 
a dimension-six Higgs operator.
This addition enables a strong first-order phase transition to develop even with 
a Higgs boson mass
well above the current direct limit of 114~GeV.  
The~$\varphi^6$ term can be generated for instance by strong dynamics at 
the TeV scale or by integrating out heavy particles like an additional singlet scalar field.
We discuss conditions to comply with electroweak
precision constraints, and point out how future experimental measurements
of the Higgs self couplings could test the idea.

\end{abstract}

\maketitle


\maketitle


\setcounter{equation}{0}

{\bf Baryogenesis and the Standard Model:} The observed large
baryon asymmetry requires natural law to
obey three principles:  baryon number violation, $C$ and $CP$ violation,
and out-of-equilibrium dynamics \cite{Sakharov:dj}.  
In the Standard Model (SM), baryon number violation
can occur through the electroweak 
sphaleron~\cite{Manton:1983nd,Klinkhamer:1984di}, 
which is a non-perturbative saddle-point
solution to the field equations attainable at high temperatures. 
These solutions allow transitions to topologically distinct $SU(2)$
vacua with differing baryon number.

$C$ is already violated in the SM as
well as $CP$, as evidenced in the Kaon and $B$-meson systems. Nevertheless, it 
has been thought~\cite{Gavela:dt} that $CP$ violation from the 
Kobayashi--Maskawa phase is too suppressed to play a 
dominant role in baryogenesis, although a recent work~\cite{Berkooz:2004kx} 
suggests a way to circumvent this common view. 
We note also that higher dimensional operators could well provide 
the desired $CP$ violation~\cite{Dine:1990fj}. 

In this letter, we focus on the last main challenge for the viability
of SM baryogenesis~\cite{Kuzmin:1985mm}: the requirement of
out-of-equilibrium dynamics.  This would be present in the SM if there was a strong
first order  EWPT. In this case, bubbles of 
the non-zero Higgs field vev 
nucleate from the symmetric vacuum and as they expand, 
particles in the plasma interact with the phase interface in 
a $CP$-violating way. The $CP$ asymmetry is converted into a baryon asymmetry 
by sphalerons in the symmetric phase in front of the bubble wall~\cite{Cohen:py}. One 
of the strongest constraints on EW baryogenesis comes from the 
requirement that baryons produced at the bubble wall are not washed 
out by sphaleron processes after they enter the broken phase. 
Imposing that sphaleron processes are sufficiently suppressed 
in the broken phase at the critical temperature leads to 
the constraint $\langle \varphi(T_c) \rangle /T_c \gtrsim 1$.  
This bound is very stable with respect to modifications of either the 
particle physics or of the cosmological evolution as 
was reviewed in~\cite{Servant:2001jh}. 
In the SM, the EWPT is first order if 
$m_H<72$~GeV~\cite{lattice} 
and
to suppress sphaleron processes in the broken phase would actually
require $m_H \lesssim 35$~GeV.
However, the current limit on the Higgs boson mass is well above that at 
$m_H>114\gev$~\cite{LEPEWWG:2003ih},
and the SM fails to be an adequate theory for baryogenesis.

As the hopes for a SM solution to baryogenesis faded other ideas have been 
pursued~\cite{Riotto:1999yt}.  One of the most promising ideas presented in
the last decade is from supersymmetry.  If the superpartner to the top
quark is lighter than about $150\gev$, a first-order EWPT can be induced  from
large-enough cubic interactions in the Higgs potential.  
This scenario
is getting a thorough test as searches for the light top superpartner are
rapidly closing the viable parameter space for this solution~\cite{MSSMBaryo}.  Recent
ideas to extend the particle spectrum 
may help resurrect electroweak baryogenesis in 
supersymmetry~\cite{Kang:2004pp}.

\bigskip
{\bf Low-scale cutoff theory:}
In this work, we focus on a single Higgs doublet model and we study how the dynamics of the EWPT can be affected by modifying the SM Higgs self-interactions. In contrast with previous approaches initiated by ref.~\cite{Anderson:1991zb}, we do not rely on large cubic Higgs interactions. Instead, we allow the possibility of a negative quartic coupling while the stability of the potential is restored by higher dimensional operators. We add a $\varphi^6$
non-renormalizable operator
to the SM potential, and show that it can 
induce a strong
first-order phase transition sufficient to drive baryogenesis~\cite{Zhang:1992fs}.  We have numerically checked that adding higher order terms in the potential suppressed by the same cutoff scale will give corrections 
of a few percent at most 
to the ratio $\langle \varphi(T_c) \rangle /T_c$ that we computed analytically 
while restricting ourselves to operators of dimension six or less.

The most general potential of degree six can be written, up to an irrelevant constant term,  as
\beq
V(\Phi)=\lambda \left( \Phi^\dagger\Phi-\frac{v^2}{2}\right)^2
 +\frac{1}{\Lambda^2}\left( \Phi^\dagger\Phi-\frac{v^2}{2}\right)^3
\label{eq:Vphi}
\eeq
where $\Phi$ is the SM electroweak Higgs doublet. At zero temperature the CP-even
scalar state can be expanded in terms of its zero-temperature vacuum
expectation value $\langle \varphi\rangle =v_0\simeq 246\gev$ and the physical Higgs boson $H$:
$\Phi=\varphi/\sqrt{2}=(H+v_0)/\sqrt{2}$.

At zero temperature we can minimize eq.~(\ref{eq:Vphi}) to find $\lambda$ 
and $v$ in
terms of physical parameters $m_H$ and $v_0$.  We find two possibilities
\begin{eqnarray*}
\begin{array}{cccc}
\hline\hline
{\rm case} & m^2_H & v^2 & \lambda \\
\hline 
\vrule height 14pt depth 5pt width 0pt
{\rm 1} & {\rm any~} m_H &
v_0^2 & \frac{m_H^2}{2v_0^2}>0 \\
\vrule height 15pt depth 11pt width 0pt
{\rm 2} & m_H^2 <\frac{3}{2} \frac{v_0^4}{\Lambda^2} &
 \hspace{2pt} v_0^2 \left( 1-\frac{2\Lambda^2m_H^2}{3v_0^4}\right) & 
-\frac{m_H^2}{2v_0^2}<0  \\ 
\hline\hline
\end{array}
\end{eqnarray*}
Note that, up to an irrelevant constant, the potential is unchanged by the parameter transformation: $\lambda \to - \lambda$ and $v^2 \to v^2 (1- 4 \Lambda^2 \lambda/(3v^2))$. So, Case 2 is actually physically equivalent to Case 1. And in the rest of the paper, we  restrict ourselves to  $\lambda>0$.
$\varphi=v_0$
is the global minimum of the potential as long as $m^2_H>v_0^4/\Lambda^2$, otherwise $\varphi=0$
is a deeper minimum.  The dynamics of the EWPT  depends only on the values of $m_H$ and $\Lambda$. It should be noted that, for the region of the parameter space where a first order EWPT occurs, the value of the quartic coupling at the origin will be negative, in contrast to the SM scenario. 

We approximate finite temperature effects by adding a thermal mass 
to the potential
$V(\varphi,T)=  c T^2 \varphi^2 /2+V(\varphi,0)$, where $c$ is generated by the quadratic terms ($T^2 m_i^2$ where $i$ denote all particles that acquire a $\varphi$-dependent mass) in the high-$T$ expansion of the one-loop thermal potential 
\beq
c=\frac{1}{16}\left(4 y_t^2+ 3 g^2 + g'^2 + 4 \frac{m_H^2}{v_0^2} 
- 12 \frac{v_0^2}{\Lambda^2}\right),
\eeq
$g$ and $g'$ are the $SU(2)_L$ and $U(1)_Y$ gauge couplings, and $y_t$ is the top Yukawa coupling.
The $T^2 m_i^2$ terms also generate a $T$-dependent contribution to the Higgs quartic coupling of the form $T^2 \varphi^4 /(4\Lambda^2)$. In the following, we have discarded this contribution to keep our analytical study simple. We have checked that it does not alter our results by more than a few percent
in the physically interesting region where a strongly first-order EWPT occurs. 
There is also a cubic Higgs interaction induced by finite temperature effects 
(crucial in supersymmetric baryogenesis)
 but it has a smaller role in our discussion, and should tend to make the EWPT
slightly stronger first-order. While perturbation theory 
is expected to break down at high temperature, its validity has been confirmed 
by lattice calculations in the regime where the EWPT is strongly first-order,
in the SM~\cite{lattice}
as well as in its supersymmetric extension~\cite{lattice2}.
 We therefore expect 
that the value of  $\langle \varphi(T_c) \rangle / T_c$ 
given by our naive tree level analysis be not too different from its actual value.

The critical temperature $T_c$ at which the minimum at $\varphi \neq0$
is degenerate with that at $\varphi=0$ is
\beq
\label{eq:Tc}
T_c^2= \frac{\Lambda^4 m_H^4+2\Lambda^2 m_H^2 v_0^4-3v_0^8}{16c\Lambda^2 v_0^4}.
\eeq
The vacuum expectation value of the Higgs field at the critical 
temperature in terms
of $m_H$, $\Lambda$ and $v_0$ is
\beq
\label{eq:vc}
\langle\varphi^2 (T_c)\rangle =v^2_c = \frac{3}{2}v_0^2
-\frac{m_H^2\Lambda^2}{2v_0^2}.
\eeq
We can see from eqs.~(\ref{eq:Tc}) and~(\ref{eq:vc}) 
that for any given $m_H$
there is an upper bound on $\Lambda$ to make sure that the phase transition is
first order (i.e. $v^2_c>0$), and there
is a lower bound on $\Lambda$ to make sure that the $T=0$ minimum 
at $\varphi\neq 0$
is a global minimum (i.e. $T_c^2>0$).  These two combine to give the important 
equation
\beq
\label{eq:Lambda-bounds}
\mathrm{max} \left( \frac{v_0^2}{m_H} , \frac{\sqrt{3} v_0^2}{\sqrt{m_H^2+2m_c^2}} \right)
<\Lambda < \sqrt{3}\frac{v_0^2}{m_H}
\eeq
where $m_c=v_0 \sqrt{(4 y_t^2 + 3 g^2 + g'^2)/8}\approx 200$~GeV. Note that the coefficient $c$ in the thermal mass is positive if and only if $\Lambda> {\sqrt{3} v_0^2}/{\sqrt{m_H^2+2m_c^2}}$. Thus when $\Lambda$ saturates the lower bound in eq.~(\ref{eq:Lambda-bounds}), the critical temperature is either
vanishing or infinite when $m_H$ is smaller or bigger than $m_c$ respectively.
At $m_H=m_c$ and $\Lambda=v_0^2/m_c$, the critical temperature is not uniquely defined but this is an artifact of our approximations. Around that point higher order terms in the thermal potential, like the $T^2 \varphi^4/(4\Lambda^2)$ terms or $T \varphi^3$ terms mentioned earlier,  will resolve the singularity.  These higher order terms will in particular give corrections to the bounds (\ref{eq:Lambda-bounds}) delineating the first order phase transition region. 

\begin{figure}[tb]
\includegraphics[width=8.6cm]{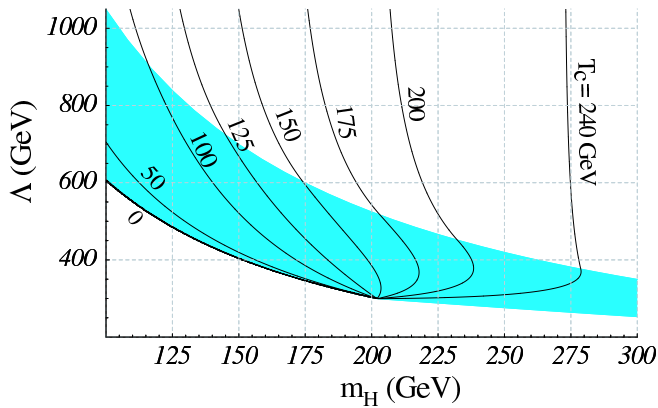}
\caption{\label{fig:Tc} 
Contours of constant $T_c$ from $0$ to $240$~GeV.  
The shaded
blue region satisfies the bounds of eq.~(\ref{eq:Lambda-bounds}). 
Above it, the EWPT is second order and the critical temperature is no 
more given by eq.~(\ref{eq:Tc}) but instead by 
$T_c^2=(2 \Lambda^2m_H^2  -3v_0^4)/4c \Lambda^2$.}
\end{figure}

\begin{figure}[tb]
\includegraphics[width=8.6cm]{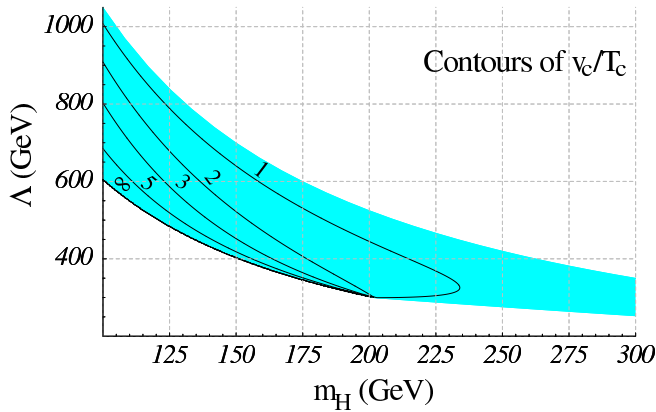}
\caption{\label{fig:vcTc} 
Contours of constant $v_c/T_c$ from $1$ to $\infty$. 
The shaded
blue region satisfies the bounds of eq.~(\ref{eq:Lambda-bounds}).}
\end{figure}

Figs.~\ref{fig:Tc} and~\ref{fig:vcTc} plot contours of constant $T_c$ and
$v_c/T_c$, respectively, in the $\Lambda$ vs.\ $m_H$ plane.  These results are encouraging and motivate a full one-loop computation of the thermal potential. Such an analysis is underway.

\bigskip
{\bf Sphaleron solution:}
We compute the sphaleron solution of this
effective field theory, eq.~(\ref{eq:Vphi}), by starting with the 
ansatz~\cite{Klinkhamer:1984di}
\begin{eqnarray}
\nonumber
W^a_i\sigma^adx^i   =  -\frac{2i}{g} f(\xi) dU\, U^{-1}, \ \
\phi  =  \frac{v_0}{\sqrt{2}} h(\xi) U
\left(\begin{array}{c} 0 \\  1\end{array}\right), 
\nonumber
\end{eqnarray}
 where $\xi=gv_0r$ and $\displaystyle U =\frac{1}{r}\left( 
\begin{array}{cc}
z & x+iy \\
-x+iy & z 
\end{array}
\right)$
and, as usual, $v_0\simeq 246\gev$.  We compute only the $SU(2)$ sphaleron
as corrections from $U(1)_Y$ are expected to be 
small~\cite{Klinkhamer:1984di,Kleihaus:1991ks}.
The functions $f$ and $h$ are solutions to two coupled nonlinear differential equations:
\bea
\nonumber
\xi^2 \frac{d^2f}{d\xi^2}& = & 2f(1-f)(1-2f)-\frac{\xi^2}{4}h^2(1-f)  \\
\frac{d}{d\xi}\left[ \xi^2\frac{dh}{d\xi}\right]& = & 2h(1-f)^2 
 +\frac{\lambda}{g^2}\xi^2(h^2-1)h \nonumber \\
& & +\frac{3}{4}\frac{v_0^2}{g^2\Lambda^2}\xi^2 h(h^2-1)^2
\nonumber
\eea
subject to the boundary conditions $f(0)=h(0)=0$ and $f(\infty)=h(\infty)=1$.
To solve these differential equations it is necessary to first expand
the solutions about their asymptotic values as $\xi\to 0$ and
$\xi\to \infty$:
\bea
\nonumber
f(\xi\to 0) & = & \xi^2/a_0^2 \ \ , \ \ h(\xi \to 0)  =  \xi/b_0\\
\nonumber
f(\xi\to \infty) & = & 1-a_\infty\exp (-\xi /2) \\
\nonumber
h(\xi\to \infty) & = & 1-(b_\infty/\xi)\exp (-\sigma \xi)
\nonumber
\eea
where $\sigma\equiv \sqrt{2\lambda/g^2}$ and $a_0, a_\infty, b_0$ 
and $b_\infty$
are constants to be determined.  

Solving for the constants $(a_0,b_0,a_\infty, b_\infty)$  is equivalent
to solving the differential equations and the boundary conditions.
We solve them by  choosing random numerical values for $a_0, a_\infty, b_0$ 
and $b_\infty$, shooting the solution from $\xi=\infty$ down to 
some intermediate
$\xi=\xi_{\rm fit}$ and also from $\xi=0$ up to $\xi=\xi_{\rm fit}$.  If both 
equations match at $\xi_{\rm fit}$ a solution has been found.  In practice,
we set up a $\chi^2$ fit to measure goodness of match, and require that
$h$, $f$ and their derivatives match to better than one part per million
before declaring that a solution has been obtained. This procedure is 
computer intensive.

\begin{figure}[tb]
\includegraphics[width=8.6cm]{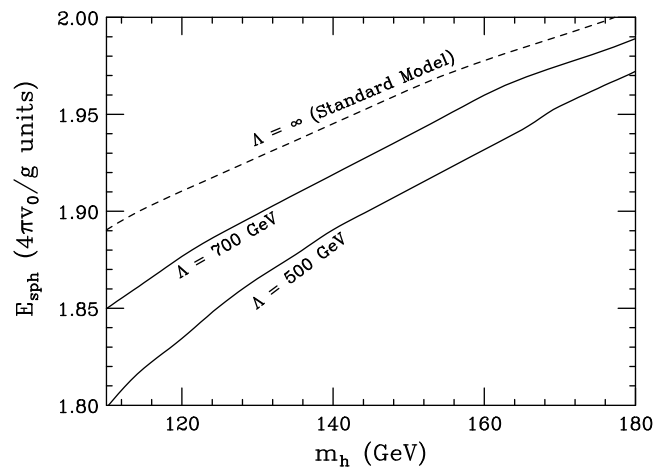}
\caption{\label{fig:Sphaleron} Sphaleron energy at zero temperature in units 
of $4 \pi v_0/g=4.75$ TeV. }
\end{figure}

After obtaining the sphaleron solution we compute the sphaleron 
energy at $T=0$ (shown in Fig.~\ref{fig:Sphaleron}) according to the equation
\bea
&&
E_{\rm sph}  =  \frac{4\pi v_0}{g}\int_0^\infty d\xi \left( 4f'^2
+\frac{8}{\xi^2}f^2(1-f)^2+\frac{1}{2}\xi^2 h'^2 +\right. \nonumber \\
& & h^2(1-f)^2+\frac{\lambda}{4g^2}\xi^2 (h^2-1)^2  +\left.\frac{v_0^2}{8g^2\Lambda^2}\xi^2 (h^2-1)^3\right).
\label{energy_sphaleron} 
\eea
It differs from the SM value by the last term, which tends to make the 
sphaleron energy slightly smaller (by only a few percent). A similar conclusion was also reached in the MSSM~\cite{Moreno:1996zm}.

The sphaleron 
energy is a crucial quantity for EW baryogenesis as the rate of baryon number 
violation in the broken phase at $T_c$ is proportional to 
$e^{-E_{\rm sph}(T_c)/T_c}$~\cite{Arnold:1987mh}. 
$E_{\rm sph}(T_c)$ is approximately given by eq.~(\ref{energy_sphaleron}) 
where $v_0$ is replaced by $v_c$, and this is how
requiring that sphaleron processes be frozen  leads to the bound 
$v_c/T_c\gtrsim 1$. Knowing whether the right-hand side of this inequality is 
1 or 1.5 is crucial in deriving the resulting bound on the Higgs mass,  and 
this depends, among other things, on the precise sphaleron energy. 
The fact that  $E_{\rm sph}$ is larger than the cutoff scale for 
a first-order phase transition is not inconsistent with the calculation of 
the rate of baryon number violation at $T_c$. Indeed,  $E_{\rm sph}$ is large 
because the sphaleron is an extended object, but its local energy density is 
always smaller than the cutoff scale. While a large amount of energy has 
to be pumped into the thermal bath to build a sphaleron configuration, this 
does not involve any local physics beyond the cutoff scale.

\bigskip
{\bf Precision electroweak constraints:}
The theory we have presented above is the SM with a low-scale
cutoff.  It is minimal in that no new particles have been introduced to
achieve the desired out-of-equilibrium first-order phase transition needed
for baryogenesis.  However, this does not mean that the phenomenology
of this model is indistinguishable from that of the SM.

The non-renormalizable operators of this theory can significantly affect
observables.  If the only additional terms are those given by
eq.~(\ref{eq:Vphi}), there would be no phenomenological constraints on this
scenario to worry about. However, a low-scale cutoff for other dimension-six
operators can be problematic for precision electroweak 
observables~\cite{Barger:2003rs}.
As an example, let us consider the following four dimension-six operators
suppressed  by the cutoff scale $\Lambda$:
\bea
\label{eq:DelL}
\Delta {\cal L} & = & \frac{\epsilon_\Phi}{\Lambda^2}
(\Phi^\dagger D_\mu \Phi)^2
+\frac{\epsilon_W}{\Lambda^2}(D_\rho W_{\mu\nu}^a)^2
+\frac{\epsilon_B}{\Lambda^2}(\partial_\rho B_{\mu\nu})^2 \nonumber \\
& & +\frac{\epsilon_F}{\Lambda^2}\bar \nu_\mu \gamma_\alpha P_L 
\mu \bar e\gamma^\alpha
P_L\nu_e.
\eea
The most sensitive precision electroweak observables are $\seff$,
$m_W$, $\Gamma_l=\Gamma(Z\to l^+l^-)$, and $\Gamma_Z$.  The percent shifts to these
observables $\Delta {\cal O}_i=\{ \seff,
m_W ({\rm GeV}),
\Gamma_l({\rm MeV}),
\Gamma_Z({\rm GeV})\}$ induced by $\Delta {\cal L}$ are
\bea
\%\left(\frac{\Delta {\cal O}}{\cal O}\right)_i =
\left(\begin{array}{rrrr}
8.57 & 6.19 & -1.47 & 4.29 \\
-4.31 & -0.55 & -0.55 & -0.65 \\
-7.20 & 1.69 & 0.93 & -3.61 \\
-7.90 & 1.00 & 1.08 & -3.93
\nonumber
\end{array}
\right)
\left(\begin{array}{c}
\tilde \epsilon_\Phi \\
\tilde\epsilon_W\\
\tilde \epsilon_B\\
\tilde\epsilon_F 
\end{array}
\right)
\eea
where $\tilde \epsilon_i=\epsilon_i\, (1\tev)^2/\Lambda^2$.
The experimental measurements of these observables 
are~\cite{LEPEWWG:2003ih}
\bea
\seff & = & 0.23150 \pm 0.00016 \nonumber\\
m_W ({\rm GeV}) & = & 80.426 \pm 0.034 \nonumber\\
\Gamma_l({\rm MeV}) & = & 83.984 \pm 0.086 \nonumber\\
\Gamma_Z({\rm GeV}) & = & 2.4952 \pm 0.0023 \nonumber
\eea
We can compare the experimental values of the observables with the 
dimension-six operator shifts induced by the cutoff scale $\Lambda$. 
The $(\Phi^\dagger D_\mu \Phi)^2$ operator appears to have the most 
substantial effect on
the precision electroweak observables.  This operator is a pure isospin
breaking operator and is equivalent to a positive shift in the $T$ parameter
in the Peskin--Takeuchi framework ($T\simeq -7.8\, \tilde \epsilon_\Phi$).

Barring some nontrivial cancellations of multiple $\epsilon_i$
contributions to the precision electroweak observables, it appears 
that $\epsilon_\Phi\lsim 10^{-2}$ is necessary if $\Lambda\lsim 1\tev$.
The other $\epsilon_i$ values are somewhat less constrained,
but likely need to be nearly as small also.  
Therefore, if this framework is to be viable there must be a small
hierarchy between the $1/\Lambda^2$ coefficient of
eq.~(\ref{eq:Vphi}) and the $\epsilon_i/\Lambda^2$ 
coefficients of eq.~(\ref{eq:DelL}). 
In the absence of a UV completion of the theory, this little hierarchy of high-dimensional operators remains unexplained.   We note in passing that
the operators can have substantially different conformal weights
if the theory at the cutoff is a strongly coupled theory where
each field gets large anomalous dimensions. Perhaps this distinguishing
property of the operators is a key to the needed hierarchy.

As a concrete example of a possible origin of the non-renormalizable Higgs self-interaction, we note that a $|\Phi|^6$ term can be generated by decoupling a massive degree of freedom. For instance, 
in a manner 
similar to ref.~\cite{Kang:2004pp} we can consider a scalar singlet $\phi_s$ coupled to the Higgs via
\begin{equation}
	\label{eq:singlet}
\Delta V= \frac{1}{2} m_s^2 \phi_s^2 + m \phi_s \Phi^\dagger \Phi + \frac{1}{2} a \phi_s^2 \Phi^\dagger \Phi.
\end{equation}
Assuming that the mass of the singlet is higher than the weak scale, integrating out this scalar degree of freedom gives rise to the additional Higgs interactions (we discard for simplicity
other terms, like $M \phi_s^3$ or $\beta \phi_s^4$, that could also be added, since
they generically will not change the conclusions):
\begin{equation}
	\label{eq:VfromS}
V_{\rm new}= - \frac{m^2}{2 m_s^2}  |\Phi|^4 + \frac{a m^2}{2 m_s^4}  |\Phi|^6 + {\cal O} \left( \frac{a^2 m^4 |\Phi|^8}{m_s^6} \right).
\end{equation}
We assume that $m$ and $m_s$ are of the same order to be able to neglect the higher-order terms in the expansion.  Therefore, if the mass scale in the singlet sector is around a TeV a $\phi^6$ term as well as a negative $\phi^4$ term are generated in the Higgs potential.
A small fine-tuning between the singlet-induced negative quartic coupling of order 1
and the initial positive quartic coupling
of the Higgs potential would be needed to produce a total quartic coupling of order
$\sim -0.1$, as is required to be sitting in the desired region of the $(m_H,\Lambda)$ plane. Meanwhile, the custodial invariant interactions of eq.~(\ref{eq:singlet}) will not lead to any of the dangerous operators eq.~(\ref{eq:DelL}).

\begin{figure}[tb]
\includegraphics[width=7.3cm]{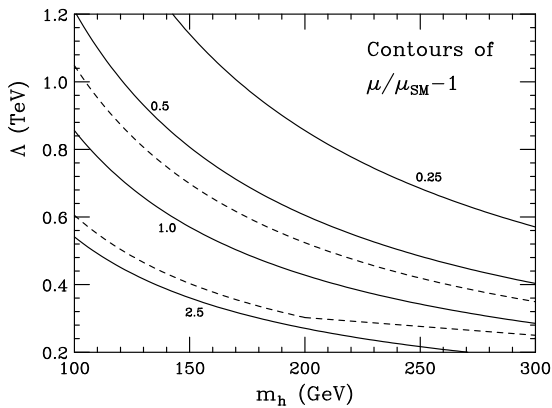}
\caption{\label{fig:cubic_coupling} Contours of constant $\mu/\mu_{\rm SM}-1$ 
in the $\Lambda$ vs.\ $m_H$ plane. 
 The dashed lines delimit the allowed region defined in 
eq.~(\ref{eq:Lambda-bounds}).}
\end{figure}

\bigskip
{\bf Higgs self-couplings as test:}
Future colliders have the opportunity to test this idea directly by
experimentally probing the Higgs potential.  When a low-scale cutoff
theory alters the Higgs potential with non-renormalizable operators, those
same operators will contribute to a shift in the Higgs self-couplings.
Expanding around the potential minimum at zero temperature we can
find the physical Higgs boson self couplings ($
{\cal L}=m_H^2 H^2 /2 + \mu H^3 / 3! + \eta H^4 / 4! +\cdots
$)
\beq
\mu =  3\frac{m_H^2}{v_0}+6\frac{v_0^3}{\Lambda^2},~~
\eta = 3\frac{m_H^2}{v_0^2}+36\frac{v_0^2}{\Lambda^2}.
\eeq
The SM couplings are recovered as $\Lambda\to \infty$.
In Fig.~\ref{fig:cubic_coupling} we plot contours of $\mu/\mu_{\rm SM}-1$ in
the $\Lambda$ vs.\ $m_H$ plane.

No experiment to date has meaningful bounds on the $H^3$ coupling.  It is estimated
that for the Higgs masses in the range needed for the first-order phase
transition presented above, a measurement of the $H^3$ coupling could be
done to within a factor of one at the LHC at $\sqrt{s}=14\tev$ with
$300\, {\rm fb}^{-1}$ integrated luminosity~\cite{Baur:2003gp}. This 
constraint or measurement
would be an interesting one for our scenario since a deviation by more than
a factor of unity is possible.  

In the more distant future, a linear collider at $\sqrt{s}=500\gev$
and $1\, {\rm ab}^{-1}$ of integrated luminosity should be able to measure
the coupling to within about $20\%$~\cite{Castanier:2001sf}, and a higher
energy linear collider, such as CLIC with $\sqrt{s}=3\tev$ and
$5\, {\rm ab}^{-1}$ integrated luminosity, should be able to measure the
self-coupling to within a few percent~\cite{Battaglia:2001nn}.  
A few-percent measurement may also
be possible at the VLHC at $\sqrt{s}=200\tev$ with $300\, {\rm fb}^{-1}$
integrated luminosity~\cite{Baur:2003gp}.

\bigskip
{\bf Conclusion:} We have shown that a strong first-order electroweak
phase transition is possible within the SM when we take into consideration
the effects of a $\varphi^6$ Higgs operator with a low cutoff.  Higgs masses
well above the 114~GeV direct limit are possible within this framework.
The main experimental test of this idea is the altered Higgs cubic self-coupling.
The LHC should be able to probe ${\cal O}(1)$ corrections, but a high-energy
linear collider will likely be required to measure the deviation at the
tens of percent level accurately.

\vspace{0.35in}

We thank J.~Cline, J.R.~Espinosa, A.~Hebecker, A.~Nelson, M.~Quir\'os and C.~Wagner for useful  comments. 
This work was supported by the Department of Energy and the Michigan Center for
Theoretical Physics. C.G. is supported in part by the RTN European Program
HPRN-CT-2000-00148
and the ACI Jeunes Chercheurs 2068. 
G.~S.~was supported in part by the US 
Department of Energy, High Energy Physics Division, under contract 
W-31-109-Eng-38 and also by the David and Lucille Packard Foundation.  

\vspace{0.35in}
\vphantom{aa}

\end{document}